\documentclass[smallextended]{svjour3}   	
\usepackage{geometry}                		
\usepackage{graphicx}				
\usepackage{amssymb}
\usepackage{amsmath}
\usepackage{amsfonts}
\usepackage{graphicx}
\usepackage{float}
\usepackage{color}
\usepackage{multicol}
\usepackage{cite}
\usepackage{tabularx}
\usepackage{array}
\newcolumntype{C}{>{$\displaystyle} c <{$}}

\title{An Unpublished Manuscript of John von Neumann on Shock Waves in Boostered Detonations} 
\subtitle{Historical Context and Mathematical Analysis}

\author{Molly Riley Knoedler \and Julianna C. Kostas \and Caroline Mary Hogan \and Harper Kerkhoff \and Chad M. Topaz}
\institute{M. R. Knoedler \at 
Department of Mathematics, University of Auckland, Auckland, New Zealand\\
\email{mkno336@aucklanduni.ac.nz}
}
\date{}
\begin{document}

\maketitle

\begin{abstract}
We report on an unpublished and previously unknown manuscript of John von Neumann and contextualize it within the development of the theory of shock waves and detonations during the nineteenth and twentieth centuries. Von Neumann studies bombs comprising a primary explosive charge along with explosive booster material. His goal is to calculate the minimal amount of booster needed to create a sustainable detonation, presumably because booster material is often more expensive and more volatile. In service of this goal, he formulates and analyzes a partial differential equation based model describing a moving shock wave at the interface of detonated and undetonated material. We provide a complete transcription of von Neumann's work and give our own accompanying explanations and analyses, including the correction of two small errors in his calculations. Today, detonations are typically modeled using a combination of experimental results and numerical simulations particular to the shape and materials of the explosive, as the complex three dimensional dynamics of detonations are analytically intractable. Although von Neumann's manuscript will not revolutionize our modern understanding of detonations, the document is a valuable historical record of the state of hydrodynamics research during and after World War II.
\end{abstract}
\small
\subsection*{Declarations}
\textbf{Funding:} We received funding from the Williams College Sciences Division and  from  National Science Foundation  grant DMS-1813752  to  Chad M. Topaz.\\
\textbf{Conflicts of interest:} We are not aware of any conflicts of interest on the part of the authors.\\
\textbf{Availability of data and materials:} Not applicable\\
\textbf{Code availability:} Not applicable\\

\section{Introduction}
The history of shock wave theory hearkens back, at least, to Sim{\'e}on-Denis Poisson's ``A Paper on the Theory of Sound,'' published in 1808 \cite{Poi1808,Poi1998,Sal2007}. In this work, Poisson, who himself built on the theories of Joseph-Louis Lagrange, used physical laws to formulate a differential equation describing the vibrations of air molecules and the resulting transmission of sound waves through the air. Poisson demonstrated that for a one dimensional spatial domain, the solution to his model consists of wave-like profiles of the air velocity $v(x)$ satisfying the implicit expression $v = f[x - (a+v)t]$, where $a$ is the speed of sound and $f$ is an arbitrary function.

In 1848, George Gabriel Stokes' ``On a Difficulty in the Theory of Sound'' picked up on the threads of Poisson's results \cite{Sto1848,Sto1998}. Stokes demonstrated that if $f$ is taken as a sine wave, $v(x)$ will develop points of infinite slope in finite time. Moreover, after that time, the solution cannot be written as an algebraic expression of the form familiar in Stokes' era. Stokes dealt with the problem of infinite slope by proposing a ``jump'' in the solution, which we now know as a shock. Stokes struggled with the idea of this new type of solution, ultimately disclosing to the reader, ``It seems to me to be of the utmost importance, in considering the application of partial differential equations to physical, and even to geometric problems, to contemplate functions apart from all ideas of algebraical expression.'' Throughout his life, Stokes vacillated on whether or not he believed his own hypothesis regarding shocks, ultimately including a disclaimer in his collected works \cite{Sal2007}. One of Stokes' principal concerns was that the mathematical model failed to preserve conservation of energy in the system.

Belief in the existence of shocks found firmer scientific footing with developments in the field of thermodynamics. In 1870, William Rankine proposed ``jump conditions'' at shocks which did, in fact, enforce conservation of energy \cite{Ran1870}. An insight that followed was that the thermodynamic process within the shock must be non-adiabatic. In 1885, Pierre-Henri Hugoniot demonstrated that in order for the system to conserve energy, there must be a jump in entropy across the shock \cite{Hug1889,Hug1998}. The two aforementioned observations led to the so-called Rankine-Hugoniot (RH) equations, which describe the conservation of mass, momentum, and energy across the shock, and which remain fundamental to understanding the properties of the discontinuity.

Our discussion thus far has drawn heavily from the history outlined in \cite{Sal2007}. The works of Poisson, Stokes, and Hugoniot that we have cited are available in a compilation of republished historical papers key to the development of shock wave theory \cite{JohChe1998}, which additionally contains relevant works of Samuel Earnshaw, Georg Friedrich Bernhard Riemann, Lord Rayleigh, Geoffrey Ingram Taylor, Hans Bethe, and Hermann Weyl.

Towards the end of the nineteenth century, the mathematical research on shock waves found application to explosions. Simple models of explosive shock waves often mention the eponymous Chapman-Jouget (CJ) state, which is the hypothesized state of the chemical products behind the reaction zone of a detonation. In this zone, products move at the minimum possible velocity greater than the speed of sound in the detonation medium in order to propagate the wave forward without interference from the shock's rarefaction waves \cite{dremin2012, guo2016prediction, cheret1999chapman}. Also known as the ``zero-reaction zone'' model, CJ theory implicitly supposes that the transformation of explosive materials into products via a shock wave is instantaneous~\cite{dremin2012}.

Proposed by Chapman in 1899 \cite{Cha1899} and built upon by Jouget in 1905-06 \cite{Jou1905,Jou1906}, the hypothesis coalesced into the first theory of detonations at the start of the twentieth century \cite{cheret1999chapman}. While the theory is useful in predicting detonation states, it is also limited in its application due to its breakdown in the case of non-ideal gases \cite{keshavarz2007}. Despite ample experimental evidence that the law was far from universal, mathematicians and physicists seeking to build generalized models of detonations during this time assumed that CJ theory was not just locally true, as was the original context of its proposal, but useful in determining the state of a flow at any point downstream of the shock \cite{cheret1999chapman}.

One of the most ambitious detonation modelers was Hungarian-American John von Neumann, a titan of scientific discovery who made numerous and influential contributions to mathematics, physics, and computer science during the first half of the twentieth century. Von Neumann's published works comprise six volumes on topics ranging from functional analysis to game theory to linear programming. During and after his involvement in the Manhattan Project, von Neumann published ten papers that use analytical and numerical methods to investigate the hydrodynamics of shock waves and detonations \cite{jvn1,jvn2,jvn3,jvn4,jvn5,jvn6,jvn7,jvn8,jvn9,jvn10}. We summarize these in Table~\ref{tab:jvnpapers}.

 \begin{table}[!htp]
  \centering
  \begin{tabularx}{\textwidth}{XcX}
  \textbf{Title} & \textbf{Year} & \textbf{Description} \\
  \hline
  Theory of Detonation Waves~\cite{jvn1} & 1942 & Analytical treatment of detonation waves in one dimension and analysis of when the CJ hypothesis cannot be assumed \\
  \hline
  Theory of Shock Waves~\cite{jvn2} & 1943 &  Analytical treatment of shocks and detonations, as well as discussion of their classifications depending on reaction types\\
  \hline
  Oblique Reflection of Shock Waves~\cite{jvn3} & 1943 & Pressure considerations based on theory and experimentation for the reflection of shocks waves colliding with oblique obstacles\\
  \hline
  Proposal and Analysis of a New Numerical Method for the Treatment of Hydrodynamical Shock Problems~\cite{jvn4} & 1944 & Numerical treatment of the differential equations governing a flow ``ignoring the possibility of shocks''\\
  \hline
  Refraction, Intersection, and Reflection of Shock Waves~\cite{jvn5} & 1945 & Interaction of shock waves in two dimensions \\
  \hline
  The Point Source Solution~\cite{jvn6} & 1947 & Analytical treatment of shock waves from a point source with infinitely high pressure in three dimensions\\
  \hline
  The Mach Effect and Height of Burst~\cite{jvn7} & 1947 & Discussion of the interference of shocks waves from a spherical detonation source reflecting on the ground\\
  \hline
  Discussion on the Existence and Uniqueness or Multiplicity of Solutions of the Aerodynamical Equations~\cite{jvn8} & 1949 & Lecture discussing the theory behind one dimensional shocks and the extreme difficulty in finding acceptable physical and mathematical principles to understand these occurances in higher dimensions \\
  \hline
  A Method for the Numerical Calculation of Hydrodynamic Shocks~\cite{jvn9} & 1949 & Numerical treatment of shocks in a one-dimensional flow with a finite reaction zone \\
  \hline
  Blast Wave Calculation~\cite{jvn10} & 1955 & Numerical treatment of a spherically symmetric point source shock wave in an ideal gas\\ \hline
  \end{tabularx}
    \caption{John von Neumann's previously known research on shock waves and detonations, listed in chronological order. Because much of this research appeared in technical reports that are difficult to obtain in original form, we cite a more accessible source, namely, the version of each paper published in a posthumous compilation of von Neumann's works \cite{jvn1963}.}
    \label{tab:jvnpapers}
  \end{table}
  
The next major leap in detonation theory was driven by necessity during World~War~II. Between 1940 and 1943, Yakov Borisovich Zel'dovich, John von Neumann and Werner D\"oring independently contributed to the development of a model of detonation that would come to be known, by their initials, as ZND theory \cite{dremin2012}. Although ZND theory uses the hypothesized CJ state, it assumes a finite (rather than infinitesimally thin) reaction zone that follows the shock wave. In this zone, materials are compressed before they undergo chemical transformation \cite{dremin2012}.

In 2017, a private collector shared with us an unpublished, handwritten manuscript of John von Neumann's \cite{manuscript}. We present a transcription and analysis of this previously unknown and unstudied work, and contextualize it within the history of shock waves and detonations as described above. The transcription is a faithful reproduction of von Neumann's writing, with exception of several commas that have been excluded for readability. 

In the manuscript \cite{manuscript}, von Neumann studies bombs comprising a primary explosive charge along with explosive booster material. His goal is to calculate the minimal amount of booster needed to create a sustainable detonation, presumably because booster material is often more expensive and more volatile. In service of this goal, he formulates and analyzes a partial differential equation based model describing a moving shock wave at the interface of detonated and undetonated material. Section \ref{sec:overview} provides an overview of the manuscript's origin and contents. In Section \ref{sec:transcription}, we present each section of \cite{manuscript} interspersed with our own analyses and explanations, including the correction of two small errors in von Neumann's work. Finally, in Section \ref{sec:conclusion}, we conclude by situating \cite{manuscript} in the context of other work of the era, and hypothesizing the motivation behind the manuscript and the possible reasons it was never published.

\section{Overview of the Manuscript}
\label{sec:overview}

\begin{figure}
\includegraphics[width=1\linewidth]{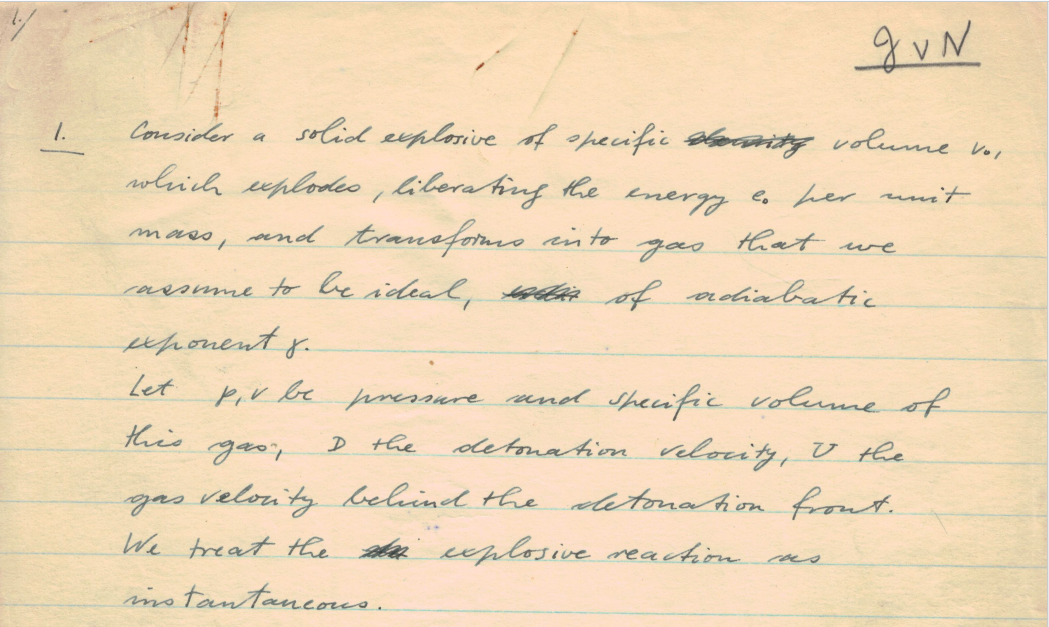}
\caption{Top portion of the first page of \cite{manuscript} in von Neumann's handwriting. Here, he begins by stating several assumptions and defining certain quantities in his detonation model. See Sec \ref{sec:transcription1} for a typed transcription.}
\label{fig:manuscript}
\end{figure}

\begin{figure}
\includegraphics[width=1\linewidth]{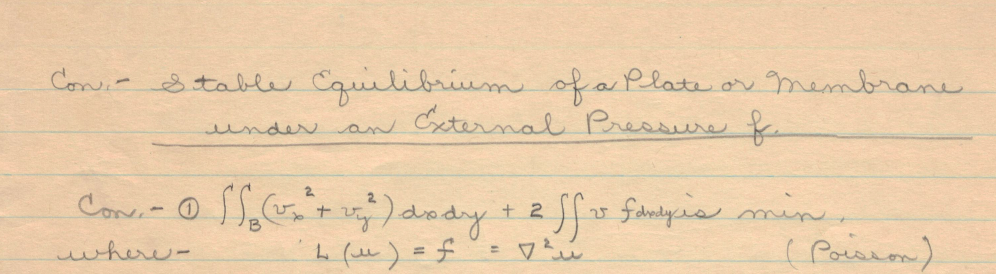}
\caption{Image of the unrelated section of the manuscript in Seeger's handwriting. We did not study this portion of the document.}
\label{fig:otherexcerpt}
\end{figure}

Von Neumann's manuscript \cite{manuscript} was obtained by a private collector through a rare book dealer, who authenticated the manuscript. The document was part of a larger archive of working manuscripts, reports, notebooks and letters produced by von Neumann and his collaborator Raymond J. Seeger, who originally compiled the archive. The manuscript we investigate, \cite{manuscript}, is not dated. The top right corner of the first page contains von Neumann's handwritten initials; see Figure~\ref{fig:manuscript}. In the pages following von Neumann's primary manuscript, there are additional pages entitled ``Stable Equilibrium of a Plate or Membrane under an External Pressure $f$'' very likely written by Seeger; see Figure \ref{fig:otherexcerpt}. We did not investigate these pages, and restrict our attention to the work confidently attributable to von Neumann. Based on the complexity of operations between steps in \cite{manuscript}, we suspect this document was a write-up of other work, rather than the original derivation.


Von Neumann aims to calculate the ratio of booster material to primary charge necessary to create an operational bomb. In this model, the more powerful (but typically more expensive and more volatile) booster material helps create a successful detonation in the primary charge. Von Neumann assumes the detonation process creates an ideal gas with uniform composition as the shock wave progresses through the solid explosive material. He further assumes that the isentropic reaction reaches thermodynamic equilibrium instantaneously with a non-adiabatic jump discontinuity at the line of the shock wave. Rather than implement a discontinuity in liberated energy between the booster and main charge, von Neumann simplifies by modeling the energy output moving away from the center of the explosive as a monotonically decreasing function. The solution von Neumann reaches is meant to describe a boostered detonation along a line, cylinder, or sphere, corresponding respectively to dimensions $q=1,2,3$.

\begin{figure}
    \centering
    \includegraphics[width=\textwidth]{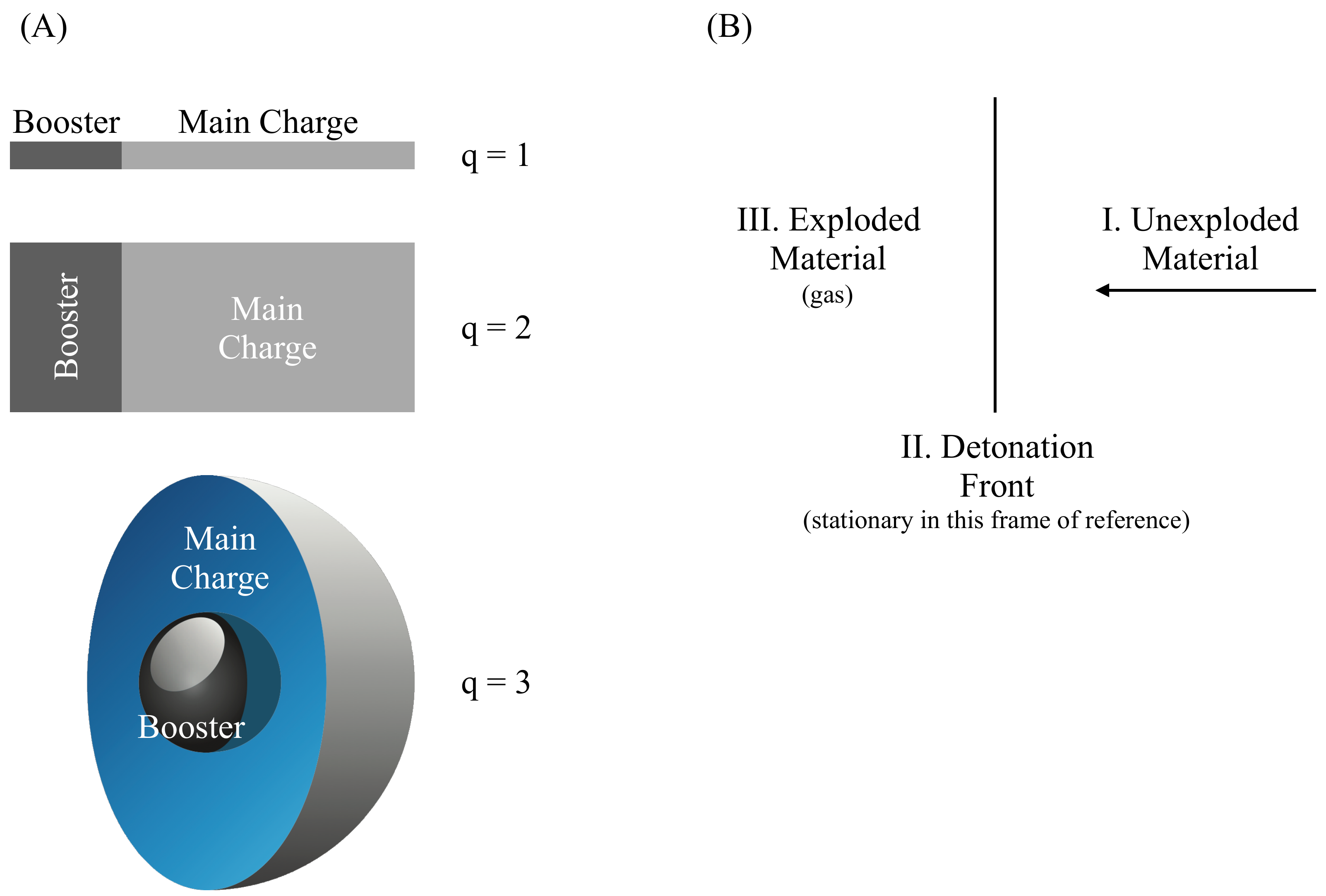}
    \caption{Explosive devices that von Neumann studies in~\cite{manuscript}. (A) Boostered explosive for a one-dimesional device ($q=1$), a cylindrical device ($q=2$), and a spherical device ($q=3$). (B) Schematic of the detonation process. In this schematic, we move into the reference frame of the detonation front. In panel (A), the front moves to the right for $q=1,2$ and radially outward for $q=3$. In (B), however, the front is stationary by design. In Region~I, the charges and, eventually, exterior gas, contact the detonation front at supersonic velocity. In Region~II, the detonation front, the material compresses and decelerates to subsonic speed. Region~III consists of exploded material, now in a gaseous phase and moving at sonic speed.}
    \label{fig:BombFigure}
\end{figure}

Table~\ref{variables} summarizes the variables and parameters used in \cite{manuscript}. Figure \ref{fig:BombFigure} provides a schematic of the boostered detonations under consideration. In the following section, we present a transcription of all 19 sections of von Neumann's manuscript \cite{manuscript}, along with our accompanying analyses. The transcription is von Neumann's wording verbatim, with the exception of erroneous commas that we have excluded to improve readability.

\begin{table}
  \centering
  \small
  \begin{tabular}{p{1cm}p{5.5cm}|p{1cm}p{5.5cm}}
  \hline
   \textbf{Symbol} & \textbf{Description} & \textbf{Symbol} & \textbf{Description}\\
  \hline
  $v_o$ & specific volume of the solid & $v$ & specific volume of the gas\\
  \hline
  $e_o$, $e$ & energy emitted per unit mass in the explosion & $p$ & pressure of the gas\\
  \hline
   $V$ & gas velocity behind the detonation front & $D$ & velocity of the detonation front\\
  \hline
  $\gamma$ & adiabatic exponent of the gas ($\gamma=\frac{c_p}{c_v}$ where $c_p$ is the specific heat capacity at constant pressure and $c_v$ is the specific heat capacity at constant volume) & $p_s$, $v_s$, $D_s$, $V_s$ & values of pressure, specific volume, detonation velocity and gas velocity behind the detonation front when the detonation velocity is at the minimum characterized by the CJ hypothesis\\
  \hline
  $w$ & the ``sluggishness'' of the explosive, a multiplier of stationary pressure $p_s$ that ensures $p$ and $D$ are greater than the stationary value; $w>1$ and $w(n)$ & $a$, $n$ & unknown values in the power law describing the energy of the explosive at point $x$; $e_0(x)=ax^{-2n}$\\
  \hline
  $q$ & dimension of the detonation; $q=$1,2 or 3, corresponding to a line, cylinder or sphere & $x$, $x'$ & position of a gas particle, where $x'<x$; initially, this can mean any $x$ but when a dependent variable of X becomes the position of a particle at $t=0$\\
  \hline
  $S_q$ & surface area of the unit sphere in $q$-dimensional space & $e_a(x)$ & average energy concentration per unit mass up to point $x$\\
  \hline
  $b$ & $\frac{e_a(x)}{e(x)}$; $b$ is the ratio of booster plus charge and $b-1$ is the amount of booster & $\bar{x}(t)$ & position of detonation front at time $t$\\
  \hline
  $k$ & $pv^{\gamma}$ & X$(x,t)$ & position at time $t$ of the gas particle that was at $x$ at $t=0$ (Lagrangian coordinate)\\
  \hline
  $A$ & for ease of notation; $\frac{2w-1}{w}\frac{1}{\gamma +1}(\frac{\gamma+w^{-1}-1}{\gamma +1})^\gamma$ & $B$ & for ease of notation; $\frac{\gamma+w^{-1}-1}{\gamma+1}$\\
  \hline
  $z$ & $\frac{x}{\bar{x}}$ & $f(z)$ & an unknown function of position; part of the solution to the Lagrangian differential equation X$_{tt}$X$_x=-vp_x$ \\
  \hline
  $C$ & undetermined multiplier of the boundary condition as $z\to \infty$ & $\mu$ & undetermined multiplier of the function $g(vz)$\\
  \hline
  $g(vz)$ & change of function from $f(z)$ & $P$ & for ease of notation; $\frac{(\gamma-1)q-2n}{(\gamma-1)q+2}$ \\
  \hline
  $s$ & change of variable from $z$ to $e^x$ & $Q$ & for ease of notation; $2\frac{(\gamma+n)q-2n}{(\gamma-1)q+2}$\\
  \hline
  $\alpha$ & for ease of notation; $\frac{\sqrt{2w-1}}{w}\frac{1}{(n+1)\sqrt{2(\gamma^2-1)a}}$ & & \\
  \hline
  \end{tabular}
  \caption{All variables and parameters in \cite{manuscript}.}
  \label{variables}
\end{table}

\section{Manuscript Transcription and Analysis}
\label{sec:transcription}
\setcounter{equation}{0}
\subsection{Sec. 1 Transcription}
\label{sec:transcription1}
\underline{1.} Consider a solid explosive of specific volume $v_0$, which explodes, liberating energy $e_0$ per unit mass and transforms it into gas that we assume to be ideal, of adiabatic exponent $\gamma$.

Let $p$, $v$ be pressure and specific volume of this gas, $D$ the detonation velocity, $V$ the gas velocity behind the detonation front. We treat the explosive reaction as instantaneous.

Then the Rankine-Hugoniot equations are:
\begin{gather}
\label{eq1} \sqrt{\frac{p}{v_0-v}} = \frac{D}{v_0} = \frac{D-V}{v} \\
\label{eq2} \frac{1}{2}p(v_0-v) = \frac{1}{\gamma-1}pv - e_0.
\end{gather}
(\ref{eq2}) is solved by
\begin{gather}
\label{eq3} v = \frac{\gamma-1}{\gamma+1}v_0 + \frac{2(\gamma-1)}{\gamma+1}e_0p^{-1}.
\end{gather}
Now (\ref{eq1}) gives
\begin{gather}
\label{eq4} D = \sqrt{\frac{\gamma+1}{2}v_0} \ \frac{p}{\sqrt{p-\frac{(\gamma-1)e_0}{v_0}}} \\
\label{eq5} V = \sqrt{\frac{2}{\gamma+1}v_0} \ \sqrt{p-\frac{(\gamma-1)e_0}{v_0}}
\end{gather}
Thus $p$ is indeterminate, but it determines $v$, $D$, $V$ by (\ref{eq3}), (\ref{eq4}), (\ref{eq5}).
\subsection{Sec. 1 Discussion}

Eq. (\ref{eq1}) is derived from the RH equation for conservation of mass and the Rayleigh line for solid explosives, and (\ref{eq2}) is derived from the RH equation for conservation of energy, also known as the Hugoniot equation \cite{le2000}.

In this section, von Neumann's derivation requires that $p=p_1-p_0$ in (\ref{eq1}) and $p=p_1+p_0$ in (\ref{eq2}), where $p_1$ is the pressure ahead of the shock wave and $p_0$ is the pressure behind. Thus, von Neumann seems to make an unstated assumption that $p_0$ is negligible. In \cite{jvn6}, he explicitly states that the difference between $p_1$ and $p_0$ is so great that $p_0$ is negligible, so he assumes $p_0$ is zero. However, the model in \cite{jvn6} assumes a gaseous point source, such that ``as the original high pressure sphere shrinks to a point, the original pressure will have to rise to infinity.'' Von Neumann does not make this assumption in any of his other published works \cite{jvn1963}, and this assumption is not made in detonation models today.

Von Neumann follows the same initial steps in \cite{manuscript} as he does in \cite{jvn6}, solving for specific volume, detonation velocity, and gas velocity. Eqs. (\ref{eq3}),(\ref{eq4}), and (\ref{eq5}) in \cite{manuscript} are analogous to (2.9$^\prime$), (2.10$^\prime$) and (2.11$^\prime$) in \cite{jvn6} (with the latter expressed here using the notation of \cite{manuscript}):

\setlength{\tabcolsep}{18pt}
\begin{equation*}
\begin{tabular}{CC}
\textbf{Expressions from \cite{manuscript}} & \textbf{Expressions from \cite{jvn6}} \\[10pt]
v=\frac{\gamma -1}{\gamma +1}v_0\ + \frac{2(\gamma-1)}{\gamma + 1}e_op^{-1} & v=\frac{\gamma -1}{\gamma +1}v_0 \\[25pt]
D = \sqrt{\frac{\gamma + 1}{2}v_0}\frac{p}{\sqrt{p-\frac{(\gamma-1)e_0}{v_0}}} & D = \sqrt{\frac{\gamma + 1}{2}v_0p} \\[25pt]
V = \sqrt{\frac{2}{\gamma + 1}v_0p}\sqrt{p-\frac{(\gamma-1)e_0}{v_0}} & V = \sqrt{\frac{2}{\gamma + 1}v_0p}
\end{tabular}
\end{equation*}
\setlength{\tabcolsep}{6pt}

These expressions differ because in \cite{manuscript}, von Neumann includes the term $e_0$ in (\ref{eq2}). He eliminates this term in \cite{jvn6} because $e_0 = p_0v_0/(\gamma - 1)$ and $p_0$ is negligible. Perhaps $e_0$ is included in \cite{manuscript} because it describes the energy of a solid, rather than a gas as in \cite{jvn6}, so gas laws do not apply.

Following this calculation, \cite{manuscript} and \cite{jvn6} diverge. The latter relies on an energetic calculation that takes into account thermal and kinetic energy, while the former uses a power law to describe energetic output; see Subsection \ref{subs47}.

\subsection{Sec. 2 - 3 Transcription}
\underline{2.} According to Chapman and Jouguet, the stationary detonation is characterised by that value of $p$, which renders $D$ a minimum. By (\ref{eq4}) this occurs when
\begin{gather*}
p = \frac{2(\gamma-1)e_0}{v_0}.
\end{gather*}
Affixing an index $s$ to all quantities referring to this state, we now have:
\begin{gather}
\label{eq6} p_s = \frac{2(\gamma-1)e_0}{v_0}, \\
\label{eq7} v_s = \frac{\gamma}{\gamma+1}v_0, \\
\label{eq8} D_s = \sqrt{2(\gamma^2-1)e_0}, \\
\label{eq9} V_s = \sqrt{\frac{2(\gamma-1)e_0}{\gamma+1}}.
\end{gather}
\underline{3.} In what follows we shall study detonation processes, in which $p$ and $D$ are higher than the Chapman-Jouguet (stationary) value. We put therefore 
\begin{gather}
\label{eq10} p = w \frac{2(\gamma-1)e_0}{v_0} = w p_s, \quad w > 1.
\end{gather}

Then (\ref{eq3}), (\ref{eq4}), (\ref{eq5}) give
\begin{gather}
\label{eq11} v = \frac{\gamma+w^{-1}-1}{\gamma+1}v_0 = \frac{\gamma+w^{-1}-1}{\gamma}v_s, \\
\label{eq12} D = \frac{w}{\sqrt{2w-1}}\sqrt{2(\gamma^2-1)e_0} = \frac{w}{\sqrt{2w-1}}D_s, \\
\label{eq13} V = \sqrt{2w-1}\sqrt{\frac{2(\gamma-1)e_0}{\gamma+1}} = \sqrt{2w-1}V_s.
\end{gather}
\subsection{Sec. 2 - 3 Discussion}
The RH equations (\ref{eq1}) and (\ref{eq2}) are underdetermined. Von Neumann uses the CJ hypothesis to find the value of pressure $p_s$ corresponding to the minimum detonation velocity $D_s$  necessary to prevent deflagration (collapse) of the detonation. The solutions for $p_s$ and $D_s$ appear in (\ref{eq6}) and (\ref{eq8}). Given these solutions, he further solves for the values of $v_s$ and $V_s$, given in (\ref{eq7}) and (\ref{eq9}). Finally, von Neumann introduces $w>1$, defined by (\ref{eq10}). Here, $w$ is a convenience of notation meant to convey that the wave propagates faster than the CJ value. Eqs. (\ref{eq10}) - (\ref{eq13}) restate the values of $p_s$, $D_s$, $v_s$, and $V_s$ using this new notation.

\subsection{Sec. 4 - 7 Transcription}
\label{subs47}
\underline{4.} Consider now an explosive of variable composition. Using the point at which the detonation begins as origin, and denoting the distance from it by $x$, we assume that the power of the explosive decreases as $x$ increases. 

The purpose of this model is to describe the mechanism of a booster. Indeed, a charge with booster consists of two zones: The inner one, $x<x_0$, being occupied by a more powerful explosive than the outer one, $x>x_0$. (These two explosives are the ``booster'' and the ``main charge''.) We replace these two zones by a continuous decrease, mainly for the sake of mathematical convenience: It will appear, that for a certain power-law of decrease the differential equations of the problem are amenable to numerical treatment. 

In order to avoid complications which are unconnected with our main problem, we assume that the density of the explosive is the same everywhere. We assume further, that the gas resulting from the explosion is always ideal, and has the same adiabatic exponent $\gamma$ everywhere. 

Thus the only quantity which can be used to express the power of the explosive, is the energy of the explosion $e_0$. Therefore $e_0$ must be a monotone decreasing function of $x$. As mentioned above, we assume a power-law:
\begin{gather}
\label{eq14} e_0 = e_0(x) = ax^{-2n}, \quad n > 0.
\end{gather}
\underline{5.} We can treat this problem in $q = 1, 2, 3$ dimensions, corresponding to plane, cylindrical, spherical detonation waves respectively. Our considerations are best carried out without specializing this $q$.\\
\underline{6.} When an explosive of this nature detonates, it is to be expected that the detonation will be more powerful at any particular place $x$, than would correspond to a homogeneous explosive, having everywhere the composition that exists at $x$. This must be so, since the explosive at $x$ is backed, and hence ``boosted'', by the more powerful explosive at $x^\prime<x$.
Hence we assume 
\begin{gather}
\label{eq15} p = wp_s, \quad w > 1.
\end{gather}
Owing to the similitude within the entire arrangement, we may expect that 
\begin{gather}
\label{eq16} w \textrm{ is independent of }x.
\end{gather}
Accordingly we make this assumption.\\
\underline{7.} In the subsequent paragraphs we will determine the stationary detonation process which is compatible with these assumptions. It will appear, that for a suitable concept of stationarity, precisely one such process exists, if $v_0$, $\gamma$, $a$, and $n$ are given. This determines $w$. 

More specifically: We shall obtain a functional relation between $n$ and $w$, which can be written as
\begin{gather}
\label{eq17} w = w(n) \textrm{ or }n = n(w). 
\end{gather}
\subsection{Sec. 4 - 7 Discussion}
Von Neumann introduces the physical description of a booster, explaining that it composes the inner layer of a two-layer explosive. For ``mathematical convenience,'' von Neumann describes the energy difference between the booster and main charge of the explosive as a monotonically decreasing function and assumes uniform density of the explosive rather than incorporating a jump discontinuity in the velocity of evolved material as the shock wave progresses between the two materials. Perhaps von Neumann never returned to this subject in his published works because this simplification compromised the usefulness of the model.

He proposes that the model describes an explosive of any dimension given that the system is constant across all angular dimensions. The generalization to several dimensions in this subsection poses some issues for the model based on previous assumptions, particularly the CJ hypothesis. Later expansions on the CJ hypothesis by Taylor and Zel'dovich showed that the hypothesis was not applicable to spheres due to an infinite gradient in the solution behind the detonation front\cite{bach1972}.

Here, von Neumann also expands on the difference between a boostered and standard detonation, and he explicitly states his goal, namely, to ``determine the stationary detonation process which is compatible with these assumptions.'' Restated, von Neumann hopes to use the more easily experimentally-determined quantities $v_o$ (specific volume of solid), $\gamma$ (adiabatic exponent of gas), $a$, and $n$ to derive $w$ (see Table~\ref{variables}), which in turn would allow him to compute the ratio of booster to charge necessary for a successful explosive, as he does later in the manuscript. Through the rest of the manuscript, von Neumann solves a Lagrangian partial differential equation for $w$.

\subsection{Sec. 8 - 9 Transcription}
\underline{8.} Before we undertake these computations, let us consider the interpretation of our procedure somewhat more closely. 

Consider the detonation at the moment when it has progressed as far as $x$. The total energy liberated up to that moment is ($S_q$ is the area of the surface of the unit sphere in $q$-dimensional space)
\begin{gather*}
\int_0^{x}{e(x') \cdot S_qx'^{q-1}}dx' = \\
= S_qa\int_o^{x}{x'^{q-2n-1}}dx'.
\end{gather*}
In order that this expression be finite, we must require $q-2n>0$, $n<\frac{q}{2}$. In order to have the outward decrease of $e(x)$, mentioned in \underline{4.}, we must require $n>0$. Hence
\begin{gather}
\label{eq18} 0<n<\frac{q}{2}.
\end{gather}
Now the above integral becomes
\begin{gather*}
\frac{S_qa}{q-2n}x^{q-2n}.
\end{gather*}
This energy is liberated in the volume
\begin{gather*}
\int_0^x{S_qx'^{q-1}}dx' = \frac{S_q}{q}x^q.
\end{gather*}
Hence the average energy concentration per unit mass is 
\begin{gather*}
e_a(x) = \frac{q}{q-2n}ax^{-2n},
\end{gather*}
and this exceeds the concentration $e(x)$ at $x$ by a factor
\begin{gather}
\label{eq19} b = \frac{e_a(x)}{e(x)} = \frac{q}{q-2n}.
\end{gather}
This factor $b$ represents in our model the ratio total (i.e. booster plus charge) : charge. Hence
\begin{gather}
\label{eq20a} \tag{20a} \textrm{ booster : charge }= b - 1 = \frac{2n}{q-2n}.
\end{gather}
\underline{9.} The formulae of \underline{7.} can now be used in this way:

An explosive may be considered ``sluggish'', if its detonation is not self-supporting. This means that the stationary detonation wave of this substance is not sufficient to initiate in it the explosive reaction in an adjacent layer. I.e. that the pressure $p_s$ or the mass velocity $V_s$---whichever may be the significant quantity---is not sufficient to produce that result. Denote the values which are actually needed to that end by $p$ and $V$. Express these quantities by (\ref{eq10}) or (\ref{eq13}), thus introducing the quantity $w>1$. This $w>1$ gives then a quantitative measure of the ``sluggishness'' of the explosive. 

Now (\ref{eq17}), (\ref{eq20a}) show how $w$ determines the relative amount of booster, which is required to detonate a given quantity of this explosive. 

These formulae express, that a given quantity of booster will only detonate a definite amount of the explosive, and no more. And this amount will also depend on the value of $q = 1, 2, 3$, i.e. on the character of the detonation wave: Plane, cylindrical, spherical.
\subsection{Sec. 8 - 9 Discussion}
Using his assumed power-law $e_0=ax^{-2n}$, von Neumann computes the amount of energy liberated in the volume of the explosion $e_0$ \emph{up to} a point $x$. This exceeds the energy liberated \emph{at} point $x$ by a factor $b$, which is the total amount of material (booster and charge) divided by the amount of charge. If the energetic output of the explosive materials and the desired dimension of the explosive, $q$, are known, then one can find $b$, the minimum necessary amount of booster material, by rearranging (\ref{eq19}) to obtain (\ref{eq20a}).

\subsection{Sec. 10 - 13 Transcription}
\underline{10.} We now pass to the computations.

Denote the time by $t$, and the position of the detonation front at that time by 
\begin{gather*}
\bar{x} = \bar{x}(t).
\end{gather*}
Then by (\ref{eq12}), (\ref{eq14})
\begin{gather*}
\frac{d\bar{x}}{dt} = D = \frac{w}{\sqrt{2w-1}}\sqrt{2(\gamma^2-1)e_0} = \\
= \frac{w}{\sqrt{2w-1}}\sqrt{2(\gamma^2-1)a} \cdot \bar{x}^{-n}.
\end{gather*}
Hence
\begin{gather*}
\bar{x}^n\frac{d\bar{x}}{dt} = \frac{w}{\sqrt{2w-1}}\sqrt{2(\gamma^2-1)a}
\end{gather*}
and (assuming $\bar{x}(0) = 0$, i.e. that the detonation began at $t = 0$), 
\begin{gather*}
\begin{cases} 
\label{eq20b} 
\tag{20b}
t = \alpha\bar{x}^{n+1} \text{ or } \bar{x} = \alpha^{-\frac{1}{n+1}}t^\frac{1}{n+1} \\
\text{with} \\
\alpha = \frac{\sqrt{2w-1}}{w} \frac{1}{(n+1)\sqrt{2(\gamma^2-1)a}}.
\end{cases}
\end{gather*}
Further (\ref{eq14}) and (\ref{eq10}), (\ref{eq11}), (\ref{eq13}) give 
\stepcounter{equation}
\begin{gather}
\label{eq21} p = w\frac{2(\gamma-1)a}{v_0}\bar{x}^{-2n}, \\
\label{eq22} v = \frac{\gamma+w^{-1}-1}{\gamma+1}v_0, \\
\label{eq23} V = \sqrt{2w-1}\sqrt{\frac{2(\gamma-1)a}{\gamma+1}}\bar{x}^{-n}.
\end{gather}
(\ref{eq21}), (\ref{eq22}) determine the adiabatic coefficient, with which the gas originates at this point of the detonation wave:
\begin{gather}
\label{eq24} k = pv^{\gamma} = w \cdot 2(\gamma-1)a \cdot (\frac{\gamma+w^{-1}-1}{\gamma+1})^{\gamma} \cdot v_0^{\gamma-1} \cdot \bar{x}^{-2n}.
\end{gather}
\underline{11.} The gas behind the detonation wave, i.e. in the interval
\begin{gather}
\label{eq25} 0<x<\bar{x} = \alpha^{-\frac{1}{n+1}}t^{\frac{1}{n+1}},
\end{gather}
is governed by the Lagrangian differential equation
\begin{gather}
\label{eq26} \textrm{X}_{tt} \textrm{X}_x = -vp_x.
\end{gather}
Here
\begin{gather}
\label{eq27} \textrm{X} = \textrm{X}(x, t) 
\end{gather}
is the position at the time $t$ of that gas particle which was at $x$ at the moment $t = 0$. $v$, $p$ obtain from the formulae 
\begin{gather}
\label{eq28} v = v_0x^{-(q-1)} \textrm{X}^{q-1} \textrm{X}_x, \\
\label{eq29} p = kv^{-\gamma},
\end{gather}
where $k = k(\bar{x})$ obtains from (\ref{eq24}), since the motion of the gas is adiabatic throughout the zone (\ref{eq25}) behind the detonation wave. 

Hence (\ref{eq26}) becomes
\begin{gather*}
\textrm{X}_{tt} = -w \cdot 2(\gamma-1)a \cdot (\frac{\gamma+w^{-1}-1}{\gamma+1})^{\gamma} \cdot \\
\cdot x^{-(q-1)} \textrm{X}^{q-1} \cdot (x^{-2n+\gamma(q-1)} \textrm{X}^{-\gamma(q-1)} \textrm{X}_{x}^{-\gamma})_x,
\end{gather*}
i.e., using (20), 
\begin{gather}
\begin{cases}
\label{eq30} 
\textrm{X}_{tt} = -A \cdot x^{-(q-1)} \textrm{X}^{q-1}(((n+1)\alpha x^n)^{-2}x^{\gamma(q-1)} \textrm{X}^{-\gamma(q-1)} \textrm{X}_{x}^{-\gamma})_x \\
\text{with} \\
A = \frac{2w-1}{w} \frac{1}{\gamma+1}(\frac{\gamma+w^{-1}-1}{\gamma+1})^{\gamma}.
\end{cases}
\end{gather}
To the differential equation (\ref{eq30}) we must add the boundary conditions. They correspond to the two ends 
\begin{gather*}
x = 0 \textrm{ and }x = \bar{x} = \alpha^{-\frac{1}{n+1}} t^{\frac{1}{n+1}}
\end{gather*}
of (\ref{eq25}). We have at 
\begin{gather} 
\label{eq31} x = 0 : \textrm{X} = 0,
\end{gather}
and at 
\begin{gather}
\begin{cases}
\label{eq32} 
x = \bar{x} = \alpha^{-\frac{1}{n+1}} t^{\frac{1}{n+1}} : \textrm{X} = x, \textrm{X}_x = B \\
\text{ with } B = \frac{\gamma+w^{-1}-1}{\gamma+1}.
\end{cases}
\end{gather}
Note, that (\ref{eq31}) and the first equation of (\ref{eq32}) express the fit with respect to position at the two ends of (\ref{eq25}), while the second equation of (\ref{eq32}) expresses the fit with respect to specific volume at the free end of (\ref{eq25}). The last obtains from (\ref{eq22}), (\ref{eq28}), remembering the $\textrm{X} = x$ at that place. The fit with respect to mass velocity at the same place could also be expressed, from (\ref{eq23}) with $V = \textrm{X}_t$. But this condition must be (and is) a consequence of the corresponding one for specific volume, since the conservation of matter has been safeguarded throughout our procedure. 

Thus the differential equation (\ref{eq30}), with boundary conditions (\ref{eq31}), (\ref{eq32}), describes the motion of the gas in (\ref{eq25}), i.e. behind the detonation wave.

\underline{12.} The similitude already referred to in (6) suggests, that the connection of 
\begin{gather*}
\frac{x}{\bar{x}} \textrm{ and }\frac{\textrm{X}}{\bar{x}}
\end{gather*}
be independent of $t$. This means, that we assume
\begin{gather}
\label{eq33} \textrm{X} = \textrm{X}(x, t) = \alpha^{-\frac{1}{n+1}} t^{\frac{1}{n+1}} f\left(\frac{x}{\alpha^{-\frac{1}{n+1}}t^{\frac{1}{n+1}}}\right) : 
\end{gather}
It is convenient to introduce 
\begin{gather}
\label{eq34} z = \frac{x}{\bar{x}} = \frac{x}{\alpha^{-\frac{1}{n+1}} t^{\frac{1}{n+1}}}.
\end{gather}
Now the differential equation (\ref{eq30}) becomes
\begin{gather}
\begin{cases}
\label{eq35}
z^2f_{zz}(z) + nzf_z(z) - nf(z) = \\
= -Az^{-(q-1)}f(z)^{q-1}(z^{-2n+\gamma(q-1)} f(z)^{-\gamma(q-1)} f_z(z)^{-\gamma})_z. 
\end{cases}
\end{gather}
And the boundary conditions (\ref{eq31}), (\ref{eq32}) become 
\begin{gather}
\label{eq36} z = 0 : f(z) = 0 \\
\label{eq37} z = 1 : f(z) = 1, f_z(z) = B. 
\end{gather}
In (\ref{eq30}), (\ref{eq32}) we expressed $A, B$ in terms of of $w$. It is now convenient to express $A, w$ in terms of $B$:
\begin{gather}
\label{eq38} A = (1-B) \cdot B^{\gamma}, \\
\label{eq39} w = \frac{1}{(\gamma+1)B - (\gamma-1)}.
\end{gather}

\underline{13.} Instead of making a direct attempt to integrate (\ref{eq35}) with (\ref{eq36}), (\ref{eq37}), we first study the conditions at (\ref{eq36}) somewhat more closely. 

Throughout (\ref{eq25}), the pressure can be expressed by (\ref{eq29}) and (\ref{eq24}), (\ref{eq28}). The same computations by which the differential equation was derived in (\ref{eq11}) give
\begin{gather*}
p = \frac{A}{v_0} ((n+1)\alpha x^n)^{-2} x^{\gamma(q-1)} \textrm{X}^{-\gamma(q-1)} \textrm{X}_x^{-\gamma}, 
\end{gather*}
hence by (\ref{eq33}), (\ref{eq34})
\begin{gather}
\label{eq40} p = \frac{A}{(n+1)^2v_0} \cdot \alpha^{\frac{2}{n+1}} t^{-\frac{2n}{n+1}} \cdot z^{-2n+\gamma(q-1)} f(z)^{-\gamma(q-1)} f_z(z)^{-\gamma}. 
\end{gather}
A simple discussion of (\ref{eq35}), which will not be given here, shows that $p \rightarrow 0$ and $p \rightarrow \infty$ are both impossible. Hence $p$ is asymptotically like $z^0$ for $z \rightarrow \infty$. By (\ref{eq40}) the same is true for $z^{-2n+\gamma(q-1)} f(z)^{-\gamma(q-1)} f_z(z)^{-\gamma}$. Hence $f(z)^{q-1}f_z(z)$ is asymptotically like $z^{q-1-\frac{2n}{\gamma}}$. Now (\ref{eq36}) necessitates $q-1-\frac{2n}{\gamma} > -1$, i.e. $n < \frac{\gamma q}{2}$, but this follows from (\ref{eq18}). Further, (\ref{eq36}) permitts to infer from the above, that $f(z)^q$ is asymptotically like $z^{q-\frac{2n}{\gamma}}$. Hence $f(z)$ is asymptotically like $z^{1-\frac{2n}{\gamma q}}$. 

So we can replace (\ref{eq36}) by this stronger requirement ($C$ undetermined but $> 0$): 
\begin{gather}
\label{eq41} f(z) = Cz^{1-\frac{2n}{\gamma q}} + ... \textrm{ for }z \rightarrow \infty.
\end{gather}

\subsection{Sec. 10 - 13 Discussion}
Here, von Neumann uses standard methods to solve a first-order in time nonlinear equation for $\bar{x}(t)$, the position of the detonation front at time $t$. Pressure is bounded, so $p \rightarrow 0$ and $p \rightarrow \infty$ are both physically unrealizable. By (35), this means that $p$ behaves asymptotically like $z^0$ for $z \rightarrow \infty$.

\subsection{Sec. 14 Transcription}
\underline{14.} It is convenient to put 
\begin{gather}
\label{eq42} f(z) = \mu g(vz). 
\end{gather}
This leaves the form of (\ref{eq35}) and (\ref{eq41}) unaffected, except that it multiplies $A$ and $C$ by 
\begin{gather*}
\mu^{-((\gamma-1)q+2)} v^{-((\gamma-1)q-2n)} \textrm{ and }\mu^{-1} v^{-(1-\frac{2n}{\gamma})}.
\end{gather*}
Thus we can choose $\mu$, $v$ so as to make both these coefficients equal to 1. Since $C$ is undetermined, we may use $v$ instead of $C$ as the undetermined quantity. In this way 
\begin{gather}
\label{eq43} \mu = A^{\frac{1}{(\gamma-1)q+2}} v^{-\frac{(\gamma-1)q-2n}{(\gamma-1)q+2}}, \\
\label{eq44} C = A^{\frac{1}{(\gamma-1)q+2}} v^{\frac{2(n+1)}{(\gamma-1)q+2} - \frac{2n}{\gamma}},
\end{gather}
ensue. 

In this way the differential equation (\ref{eq35}) becomes 
\begin{gather}
\begin{cases}
\label{eq45}
z^2g_{zz}(z) + nzg_z(z) - ng(z) = \\
= -z^{-(q-1)} g(z)^{q-1} (z^{-2n+\gamma(q-1)} g(z)^{-\gamma(q-1)} g_z(z)^{-\gamma})_z,
\end{cases}
\end{gather}
and the boundary condition (\ref{eq36}), in its stronger form (\ref{eq41}), becomes 
\begin{gather}
\label{eq46} g(z) = z^{1-\frac{2n}{\gamma q}} + ... \quad \textrm{ for }z \rightarrow 0. 
\end{gather}
The remaining boundary condition (\ref{eq37}) becomes now:
\begin{gather}
\begin{cases}
\label{eq47}
z = v: \quad g(z) = A^{-\frac{1}{(\gamma-1)q+2}} \cdot v^{\frac{(\gamma-1)q-2n}{(\gamma-1)q+2}}, \\
\quad \quad \quad \quad g_z(z) = A^{-\frac{1}{(\gamma-1)q+2}}B \cdot v^{-\frac{2(n+1)}{(\gamma-1)q+2}}.
\end{cases}
\end{gather}
Since $v$ is arbitrary, we can formulate (\ref{eq37}) like this: There must exist a $z$ for which 
\begin{gather}
\label{eq48} g(z) = A^{-\frac{1}{(\gamma-1)q+2}} \cdot z^{\frac{(\gamma-1)q-2n}{(\gamma-1)q+2}}, \\
\label{eq49} g_z(z) = A^{-\frac{1}{(\gamma-1)q+2}}B \cdot z^{-\frac{2(n+1)}{(\gamma-1)q+2}}.
\end{gather}
This $z$ then determines $v$ by 
\begin{gather}
\label{eq50} v = z. 
\end{gather}
We formulate (\ref{eq48}), (\ref{eq49}): 
\begin{gather}
\label{eq51} B = \frac{zg_z(z)}{g(z)}, \\
\label{eq52} A = z^{(\gamma-1)q-2n} \cdot g(z)^{-((\gamma-1)q+2)}. 
\end{gather}
As we saw at the end of (12), $B$ is an undetermined quantity, too. So we can interpret (\ref{eq51}) as defining $B$, and eliminate $A$, $B$ from (\ref{eq51}), (\ref{eq52}) by means of (\ref{eq38}). In this way the condition 
\begin{gather}
\begin{cases}
\label{eq53}
z^{-(\gamma-1)(q-1)+(2n+1)} \cdot g(z)^{(\gamma-1)(q-1)} \cdot g_z(z)^{\gamma} \cdot \\
\quad \quad \cdot (g(z) - zg_z(z)) = 1
\end{cases}
\end{gather}
obtains. 

We can also substitute (\ref{eq51}) in (\ref{eq39}). This gives
\begin{gather}
\label{eq54} w = \frac{g(z)}{(\gamma+1)zg_z(z)-(\gamma-1)g(z)}. 
\end{gather}

\subsection{Sec. 14 Discussion}
The manuscript's multipliers of $A$ and $C$ appear to contain a minor error. When we substitute $\mu g(vz)$ into (35), we obtain
\[z^2\mu v^2g_{zz}(vz)+nz\mu g_z(vz)-n\mu g(vz)\]
\[=-Az^{-(q-1)}\mu^{q-1}g(vz)^{q-1}(z^{-2n+\gamma(q-1)}\mu ^{-\gamma(q-1)}g(vz)\mu^{-\gamma}v^{-\gamma}g_z(vz)^{-\gamma})_z\]
Substituting in $\frac{z}{v}$ for $z$, we obtain
\[\frac{z}{v}^2\mu v^2g_{zz}(z)+n\frac{z}{v}\mu g_z(z)-n\mu g(z)\]=
\[-A\frac{z}{v}^{-(q-1)}\mu^{q-1}g(z)^{q-1}(\frac{z}{v}^{-2n+\gamma(q-1)}\mu ^{-\gamma(q-1)}g(z)\mu^{-\gamma}v^{-\gamma}g_z(z)^{-\gamma})_{\frac{z}{v}}\]
Pulling $u$ and $v$ terms out does then yield the multiplier of $A$ given in the manuscript, leaving the form of (\ref{eq35}) unaffected.

However, when we substitute $\mu g(vz)$ into (41), we obtain
\[\mu g(vz)=Cz^{1-\frac{2n}{\gamma q}}\]
\[g(vz)=\mu^{-1}Cz^{1-\frac{2n}{\gamma q}}.\]
Again, substitute in $\frac{z}{v}$ for $z$; then
\[g(z)=\mu^{-1}C\frac{z}{v}^{1-\frac{2n}{\gamma q}}\]
\[g(z)=\mu^{-1}v^{-(1-\frac{2n}{\gamma q})}Cz^{1-\frac{2n}{\gamma q}}.\]
The multiplier given by von Neumann lacks the $q$ in the power of $v$; the $q$ is again missing in (44). However, the boundary condition given in (46) is unaffected by this minor error. The rest of the section follows as stated by von Neumann.

\subsection{Sec. 15 - 17 Transcription}
\underline{15.} Summing up:

After $\gamma$, $q$ and $n$ are chosen, the differential equation (\ref{eq45}) must be integrated, beginning at $z = 0$ with (\ref{eq46}). The solution must be continued up to the point $z$ where (\ref{eq53}) holds. This $z$ then determines $w$ by (\ref{eq54}). This is the process by which (\ref{eq17}) is obtained. 

\underline{16.} The process of integrating (\ref{eq45}) can be simplified by putting
\begin{gather}
\label{eq55} z = e^s \\
\label{eq56} g(z) = e ^{Ps} \mu(s)
\end{gather}
where
\begin{gather}
\label{eq57} P = \frac{(\gamma-1)q-2n}{(\gamma-1)q+2}. 
\end{gather}
Then (\ref{eq45}) becomes
\begin{gather}
\begin{cases}
\label{eq58}
(\frac{d}{ds} + (P-1))(\frac{d}{ds} + P)\mu(s) + n(\frac{d}{ds} + P)\mu(s) - n\mu(s) = \\
= -\mu(s)^{q-1}(\frac{d}{ds} + Q)(\mu(s)^{-\gamma(q-1)}((\frac{d}{ds} + P)\mu(s))^{-\gamma}), 
\end{cases}
\end{gather}
where
\begin{gather}
\label{eq59} Q = -2n + \gamma q(1-P) = 2\frac{(\gamma+n)q - 2n}{(\gamma-1)q + 2}. 
\end{gather}
Now put
\begin{gather}
\label{eq60} w(s) = (\frac{d}{ds} + P)\mu(s), 
\end{gather}
so that
\begin{gather}
\label{eq61} g_z(z) = e^{(P-1)s} w(s). 
\end{gather}
Then
\begin{gather}
\label{eq62} \frac{d}{ds} = \frac{d\mu}{ds} \frac{d}{d\mu} = (w-P\mu)\frac{d}{d\mu}, 
\end{gather}
and so (\ref{eq58}) becomes
\begin{gather*}
((w - P\mu)\frac{d}{d\mu} + (P-1))w + nw - n\mu = \\
\quad = -\mu^{q-1}((w-P\mu)\frac{d}{d\mu} + Q)(\mu^{-\gamma(q-1)}w^{-\gamma}),
\end{gather*}
i.e.
\begin{gather}
\begin{cases}
\label{eq63}
\frac{dw}{d\mu} = \\
= \frac{1}{w-P\mu} \frac{[(1-P-n)w+n\mu] + [\gamma(q-1)w-(\gamma(q-1)P+Q)\mu]\mu^{-((\gamma-1)(q-1)+1)} w^{-\gamma}}{1 - \gamma \mu^{-((\gamma-1)(q-1)+1)} w^{-(\gamma+1)}}.
\end{cases}
\end{gather}

\underline{17.} (\ref{eq62}) gives further
\begin{gather}
\label{eq64} s = \int{\frac{d\mu}{w-P\mu} + \textrm{ Constant }}, 
\end{gather}
and $z \rightarrow 0$ clearly means $s \rightarrow -\infty$ by \ref{eq55}. (\ref{eq46}) becomes, by (\ref{eq55}), (\ref{eq56})
\begin{gather}
\label{eq65} \mu(s) = e^{\frac{Q}{\gamma q}s} + . . . \quad \textrm { for }s \rightarrow -\infty.
\end{gather}
Similarly, differentiation of (\ref{eq46}) gives, by (\ref{eq55}), (\ref{eq60})
\begin{gather}
\label{eq66} w(s) = (1-\frac{2n}{\gamma q}) e^{\frac{Q}{\gamma q} s} + . . . \quad \textrm { for }s \rightarrow \infty.  
\end{gather}
Before we discuss (\ref{eq65}), (\ref{eq66}) any further, let us also reformulate (\ref{eq53}) and (\ref{eq54}). (\ref{eq53}) becomes
\begin{gather*}
\mu^{(\gamma-1)(q-1)} \cdot w^{\gamma} \cdot (\mu-w) = 1, 
\end{gather*}
i.e.
\begin{gather}
\label{eq67} \mu - w = \mu^{-(\gamma-1)(q-1)} w^{-\gamma}. 
\end{gather}
(\ref{eq54}) becomes
\begin{gather}
\label{eq68} w = \frac{\mu}{(\gamma+1)w - (\gamma-1)\mu}.
\end{gather}
\subsection{Sec. 15 - 17 Discussion}
The goal of these sections is to solve for the Lagrangian coordinate $X(x,t)$, which describes the position at time $t$ of a gas particle that was initially at position $x$. Von Neumman obtains the solution through a series of variable transformations, substitutions, asymptotic arguments, and integrations.

\subsection{Sec. 18 - 19 Transcription}
\underline{18.} Let us now return to (\ref{eq65}), (\ref{eq66}). 

By (\ref{eq18}) the numerator of $Q$ in (\ref{eq59}) is
\begin{gather*}
(\gamma+n)q-2n = \gamma q + (q-2)n \begin{cases} \geq \gamma q > 0 \\
\text{ for }q = 1, 2, \\
\end{cases}
\begin{cases} \\ 
= 3\gamma - n \geq 3\gamma - \frac{3}{2} > \\
> 0 \\
\text{ for }q = 3. 
\end{cases}
\end{gather*}
Consequently in any event 
\begin{gather}
\label{eq69} Q > 0.
\end{gather}
Thus (\ref{eq65}), (\ref{eq66}) imply that on that boundary point 
\begin{gather}
\label{eq70} u, w \rightarrow 0, \frac{w}{u} \rightarrow 1 - \frac{2n}{\gamma q}. 
\end{gather}
Conversely: Combining (\ref{eq70}) with (\ref{eq64}) gives 
\begin{gather*}
s \sim \int{\frac{du}{(1-\frac{2u}{\gamma q}-P)u}} = \int{\frac{du}{\frac{Q}{\gamma q} u}} = \\
\quad = \frac{\gamma q}{Q} \ln{u} + \textrm{ Constant }. 
\end{gather*}
We can adjust this constant so that 
\begin{gather*} 
s \sim \frac{\gamma q}{Q} \ln{u}
\end{gather*}
results, hence $s \rightarrow -\infty$, $u \sim e^{\frac{Q}{\gamma q} s}$, and so by (\ref{eq70}) $w \sim (1 - \frac{2n}{\gamma q})e^{\frac{Q}{\gamma q}s}$. Thus (\ref{eq65}), (\ref{eq66}) are valid. 

In other words: (\ref{eq65}), (\ref{eq66}) may be replaced by (\ref{eq70}).

\underline{19.} Summing up: (Cf. (15)) 

After $\gamma$, $q$ and $n$ are chosen, the differential equation (\ref{eq63}) must be integrated, beginning with (\ref{eq70}). 

The solution must be continued up to the point $u$, $w$ where (\ref{eq67}) holds. These $u$, $w$ then determine $w$ by (\ref{eq68}). 

This is the process by which (\ref{eq17}) is obtained.

\subsection{Sec. 18 - 19 Discussion}
\indent Although $Q>0$ in (59), von Neumann's statement in the unnumbered equation between (68) and (69) appears incorrect. Our own analysis is as follows.

When $q=1$, the numerator of $Q$ is greatest when $n\to 0$ and smallest when $n\to q/2$. In fact, $Q$ can be \textit{less than} 0 for $\gamma<1/2$; however, the quantity $\gamma$ is always greater than 1 by definition. Recall that $\gamma$ is the ratio of the heat capacity of an ideal gas at constant pressure to the heat capacity at constant volume. This quantity can also be expressed as $1+2/f$, where $f$ is the degrees of freedom of a molecule, a positive value. Hence $\gamma > 1$, and $Q<0$ is unrealizable. When $q=2$, the numerator of $Q$ is always $\gamma q$, and $Q>0$ for all $n$. When $q=3$, the numerator of $Q$ is smallest when $n\to 0$ and greatest when $n\to q/2$. Similarly to the case when $q=1$, $Q$ can be less than zero for $\gamma<1/2$. \par
In these final sections, von Neumann shows that the differential equation $dw/d\mu$ found in (\ref{eq63}) can be integrated, and by the inequalities above,
\[
w \sim \left(1 - \frac{2n}{\gamma q}\right)e^{\frac{Q}{\gamma q}s}.
\]
Von Neumann's goal was to compute the functional relationship between $w$ and $n$, and he succeeds with the expression above.

\section{Discussion and Conclusions}
\label{sec:conclusion}
In his published work on shock waves \cite{jvn1} that ultimately led to ZND theory, von Neumann shows ``when the so-called Chapman-Jouguet hypothesis is true, and what formulae are to be used when it is not true.''  He states later in \cite{jvn1} that ``it is hoped that this will  connect  the  present  theory  with  the  difficult  questions  of  initiating  a  detonation  of primers and boosters,'' and in \cite{jvn2} he states that he will publish a subsequent report on boosters.

We could not locate any such work specific to boosters. As for the handwritten manuscript \cite{manuscript}, it is difficult to place the exact time of its writing. On the one hand, we might assume that it is the paper on boosters von Neumann hinted at in 1942 and 1943 \cite{jvn1,jvn2} but never published, and that he was working on it simultaneously or immediately following these works. On the other hand, \cite{jvn1,jvn2} take into account important aspects of detonations that von Neumann does not include in \cite{manuscript}, perhaps suggesting that \cite{manuscript} came first. For example, the 1943 publication \cite{jvn2} specifically states the differences in modelling that must be accounted for in one, two or three dimensions; these are not addressed in \cite{manuscript}. Additionally, although von Neumann states repeatedly in \cite{manuscript} that the CJ hypothesis is generally true, by 1942 he had already shown when it failed as a predictive method. He is careful to note this failure in subsequent works \cite{jvn1963}, and yet no mention of it is made in \cite{manuscript}. Moreover, the majority of von Neumann's works on the theory of shock waves following \cite{jvn1,jvn2} (excluding those concerned with shock wave reflection) have the following characteristics:
\begin{itemize}
    \item incorporation of thermodynamic information that is absent in \cite{manuscript},
    \item substantial usage of numerical methods rather than analytical ones,
    \item formulation for detonation in a particular dimension, and
    \item points or homogeneous spheres as the source of detonation.
\end{itemize}
In most of his shock wave papers, von Neumann emphasizes the infancy of the theory and the difficulty in finding theories that are generalizable and physically justifiable. The simplicity of the models in his published works perhaps suggests why von Neumann never published the more ambitious manuscript we have analyzed here. While von Neumann's step-by-step derivations in the unpublished manuscript are mostly accurate, several fundamental assumptions of the model cannot be justified and the nuance of his discussion on the topic in his published works makes clear that he was aware of these difficulties. For example, his liberal use of the CJ hypothesis, the simplification of the discontinuity between the booster and the main charge in terms of energetic output to a monotonic decreasing function, the lack of specific thermodynamic laws in the energetic output equations, the generalization to multiple dimensions, and the exclusion of $p_0$ without explanation are much less sophisticated than the work done in his published papers.

Von Neumann was diagnosed with cancer in 1955 and subsequently placed under security for fear that he would ``reveal military secrets while heavily medicated,'' passing away in 1957 \cite{macrae2019john}. To date, there is no generalized theory regarding boostered detonations like the one von Neumann hoped to create in \cite{manuscript}. In the 1960s, experimental observation made clear that the one dimensional theories could not, in general, capture the complex structures of three dimensional detonations. Detonations have asymmetrical dynamics specific to the material used. Those materials and their chemical properties result in complications that prohibit the creation of a generalizable theory. Instead, investigations typically came to rely on experiments and numerical simulations. From that point of view, the handwritten manuscript is not an undiscovered model that will propel forward the modern understanding of explosive shock waves; however, it is an intriguing analysis performed by a mathematical powerhouse. The manuscript \cite{manuscript} sheds light on the scientific hurdles faced in the development of detonation theory during a time of pressure to understand explosives as they were being rapidly developed and produced during World War II.

\begin{acknowledgements}
We are grateful to the manuscript owner, Derek Kaufman, for  making von Neumann's work available for research, for helping us conceive this project, and for providing input based on a careful reading of this article. We received funding from the Williams College Sciences Division and from National Science Foundation grant DMS-1813752 to CMT. Finally, we would like to thank John von Neumann for leaving this intriguing document behind.
\end{acknowledgements}

\noindent On behalf of all authors, the corresponding author states that there is no conflict of interest.
\bibliographystyle{spmpsci} 
\bibliography{JVNBib}

\begin{thebibliography}{10}
\providecommand{\url}[1]{{#1}}
\providecommand{\urlprefix}{URL }
\expandafter\ifx\csname urlstyle\endcsname\relax
  \providecommand{\doi}[1]{DOI~\discretionary{}{}{}#1}\else
  \providecommand{\doi}{DOI~\discretionary{}{}{}\begingroup
  \urlstyle{rm}\Url}\fi

\bibitem{bach1972}
Bach, G., Lee, J.: On the propagation of spherical detonation waves.
\newblock Tech. rep., McGill Unniversity, Dept. of Mechanical Engineering
  (1972)

\bibitem{Cha1899}
Chapman, D.: On the rate of explosion in gases.
\newblock Phil. Mag. Ser. 5 \textbf{47}(284), 90--104 (1899)

\bibitem{cheret1999chapman}
Ch{\'e}ret, R.: {C}hapman-{J}ouguet hypothesis 1899-1999: One century between
  myth and reality.
\newblock Shock Waves \textbf{9}(5), 295--299 (1999)

\bibitem{dremin2012}
Dremin, A.: Toward Detonation Theory.
\newblock Springer (2012)

\bibitem{guo2016prediction}
Guo, D., Zybin, S., An, Q., Goddard, W., Huang, F.: Prediction of the
  {C}hapman-{J}ouguet chemical equilibrium state in a detonation wave from
  first principles based reactive molecular dynamics.
\newblock Phys. Chem. Chem. Phys. \textbf{18}(3), 2015--2022 (2016)

\bibitem{Hug1889}
Hugoniot, H.: Propagation du mouvement dans les corps et plus sp{\'e}cialement
  dans les gaz parfaits.
\newblock J. \'Ecole Polytech. \textbf{57}, 1--125 (1889)

\bibitem{Hug1998}
Hugoniot, H.: On the propagation of motion in bodies and in perfect gases in
  particular - ii.
\newblock In: J.N. Johnson, R.~Ch{\'e}ret (eds.) Classic Papers in Shock
  Compression Science, chap.~9, pp. 245--358. Springer (1998)

\bibitem{JohChe1998}
Johnson, J., Ch{\'e}ret, R.: Classic Papers in Shock Compression Science.
\newblock Springer (1998)

\bibitem{Jou1905}
Jouguet, E.: Sur la propagation des r{\'e}actions chimiques dans les gaz, pt.
  1.
\newblock J. Maths. Pure Appl. Ser. 6 \textbf{60}(4), 347--425 (1905)

\bibitem{Jou1906}
Jouguet, E.: Sur la propagation des r{\'e}actions chimiques dans les gaz, pt.
  2.
\newblock J. Maths. Pure Appl. Ser. 6 \textbf{61}(1), 5--86 (1906)

\bibitem{keshavarz2007}
Keshavarz, M., Warey, P.: New Research on Hazardous Materials.
\newblock Nova Science (2007)

\bibitem{le2000}
Le~Roy, F.: General laws for propagation of shock waves through matter.
\newblock Handbook of Shock Waves p. 143 (2000)

\bibitem{macrae2019john}
Macrae, N.: John von Neumann: The scientific genius who pioneered the modern
  computer, game theory, nuclear deterrence, and much more.
\newblock Plunkett Lake Press (2019)

\bibitem{manuscript}
von Neumann, J.: Unpublished manuscript

\bibitem{jvn10}
von Neumann, J.: Blast wave calculation (1955).
\newblock In: Collected Works, Vol. VI: Theory of Games, Astrophysics,
  Hydrodynamics and Meteorology. Pergamon (1963)

\bibitem{jvn1963}
von Neumann, J.: Collected Works, Vol. VI: Theory of Games, Astrophysics,
  Hydrodynamics and Meteorology.
\newblock Pergamon (1963)

\bibitem{jvn8}
von Neumann, J.: Discussion on the existence and uniqueness or multiplicity of
  solutions of the aerodynamical equations (1949).
\newblock In: Collected Works, Vol. VI: Theory of Games, Astrophysics,
  Hydrodynamics and Meteorology. Pergamon (1963)

\bibitem{jvn7}
von Neumann, J.: The {M}ach effect and height of burst (1947).
\newblock In: Collected Works, Vol. VI: Theory of Games, Astrophysics,
  Hydrodynamics and Meteorology. Pergamon (1963)

\bibitem{jvn9}
von Neumann, J.: A method for the numerical calculation of hydrodynamic shocks
  (1949).
\newblock In: Collected Works, Vol. VI: Theory of Games, Astrophysics,
  Hydrodynamics and Meteorology. Pergamon (1963)

\bibitem{jvn3}
von Neumann, J.: Oblique reflection of shock waves (1943).
\newblock In: Collected Works, Vol. VI: Theory of Games, Astrophysics,
  Hydrodynamics and Meteorology. Pergamon (1963)

\bibitem{jvn6}
von Neumann, J.: The point source solution (1947).
\newblock In: Collected Works, Vol. VI: Theory of Games, Astrophysics,
  Hydrodynamics and Meteorology. Pergamon (1963)

\bibitem{jvn4}
von Neumann, J.: Proposal and analysis of a new numerical method for the
  treatment of hydrodynamical shock problems (1944).
\newblock In: Collected Works, Vol. VI: Theory of Games, Astrophysics,
  Hydrodynamics and Meteorology. Pergamon (1963)

\bibitem{jvn5}
von Neumann, J.: Refraction, intersection, and reflection of shock waves
  (1945).
\newblock In: Collected Works, Vol. VI: Theory of Games, Astrophysics,
  Hydrodynamics and Meteorology. Pergamon (1963)

\bibitem{jvn1}
von Neumann, J.: Theory of detonation waves (1942).
\newblock In: Collected Works, Vol. VI: Theory of Games, Astrophysics,
  Hydrodynamics and Meteorology. Pergamon (1963)

\bibitem{jvn2}
von Neumann, J.: Theory of shock waves (1943).
\newblock In: Collected Works, Vol. VI: Theory of Games, Astrophysics,
  Hydrodynamics and Meteorology. Pergamon (1963)

\bibitem{Poi1808}
Poisson, S.: M\'emoire sur la th\'eorie du son.
\newblock J. \'Ecole Polytech. \textbf{14}(7), 319--392 (1808)

\bibitem{Poi1998}
Poisson, S.: A paper on the theory of sound.
\newblock In: J.N. Johnson, R.~Ch{\'e}ret (eds.) Classic Papers in Shock
  Compression Science, chap.~1, pp. 3--65. Springer (1998)

\bibitem{Ran1870}
Rankine, W.: On the thermodynamic theory of waves of finite longitudinal
  disturbance.
\newblock Phil. Trans. Roy. Soc. Lond. \textbf{160}, 277--288 (1870)

\bibitem{Sal2007}
Salas, M.: The curious events leading to the theory of shock waves.
\newblock Shock Waves \textbf{16}(6), 477--487 (2007)

\bibitem{Sto1848}
Stokes, G.: On a difficulty in the theory of sound.
\newblock Phil. Mag. \textbf{33}, 349--356 (1848)

\bibitem{Sto1998}
Stokes, G.: On a difficulty in the theory of sound.
\newblock In: J.N. Johnson, R.~Ch{\'e}ret (eds.) Classic Papers in Shock
  Compression Science, chap.~2, pp. 71--79. Springer (1998)

\end{thebibliography}

\end{document}